\documentclass{IEEEoj}
\usepackage{cite}
\usepackage{amsmath,amssymb,amsfonts,amsthm}
\usepackage{algorithm}
\usepackage{algpseudocode}
\usepackage{graphicx,color}
\usepackage{textcomp}
\usepackage{mathtools}
\usepackage[caption=false, font=footnotesize]{subfig}
\usepackage[colorlinks = true, linkcolor = blue, urlcolor=blue, citecolor =
red, anchorcolor = green,]{hyperref}

\def\BibTeX{{\rm B\kern-.05em{\sc i\kern-.025em b}\kern-.08em
    T\kern-.1667em\lower.7ex\hbox{E}\kern-.125emX}}
\AtBeginDocument{\definecolor{ojcolor}{cmyk}{0.93,0.59,0.15,0.02}}
\def\OJlogo{\vspace{-4pt}\hskip-4pt\includegraphics[height=18pt]{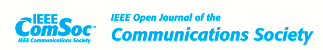}}

\begin{document}
\receiveddate{XX Month, XXXX}
\reviseddate{XX Month, XXXX}
\accepteddate{XX Month, XXXX}
\publisheddate{XX Month, XXXX}
\currentdate{\today}
\doiinfo{OJCOMS.2024.011100}

\title{Quantum Switches for Gottesman-Kitaev-Preskill Qubit-based All-Photonic Quantum Networks}

\author{MOHADESEH AZARI\IEEEauthorrefmark{1} 
, PAUL POLAKOS \IEEEauthorrefmark{2}, AND KAUSHIK P. SESHADREESAN\IEEEauthorrefmark{1}
}
\affil{Department of Informatics and Networked Systems, University of Pittsburgh, Pittsburgh, PA 15260, USA}
\affil{Cisco Systems, New York, NY 10119, USA}
\authornote{This work was supported by NSF grant 2204985 and Cisco Systems.}

\begin{abstract}
The Gottesman-Kitaev-Preskill (GKP) code, being information theoretically near optimal for quantum communication over Gaussian thermal-loss optical channels, is likely to be the encoding of choice for advanced quantum networks of the future. 
Quantum repeaters based on GKP-encoded light have been shown to support high end-to-end entanglement rates across large distances despite realistic finite squeezing in GKP code preparation and homodyne detection inefficiencies.  
Here, we introduce a quantum switch for GKP-qubit-based quantum networks, whose architecture involves multiplexed GKP-qubit-based entanglement link generation with clients, and their all-photonic storage, together enabled by GKP-qubit graph state resources.  
For bipartite entanglement distribution between clients via entanglement swapping, the switch uses a multi-client generalization of a recently introduced \textit{entanglement-ranking-based link matching} protocol heuristic.
Since generating the GKP-qubit graph state resource is hardware intensive, given a total resource budget and an arbitrary layout of clients, we address the question of their optimal allocation towards the different client-pair connections served by the switch such that the sum throughput of the switch is maximized while also being fair in terms of the individual entanglement rates. 
We illustrate our results for an exemplary data center network, where the data center is a client of a switch and all of its other clients aim to connect to the data center alone---a scenario that also captures the general case of a gateway router connecting a local area network to a global network. 
Together with compatible quantum repeaters, our quantum switch provides a way to realize quantum networks of arbitrary topology.
\end{abstract}


\maketitle

\section{INTRODUCTION}
Quantum networking is at the heart of the ongoing second quantum revolution~\cite{DM03}. 
At small distance scales, modular architectures for quantum computers comprised of extensive collections of interconnected small, finite-sized quantum logic units provide a way to scale up quantum computing power~\cite{QuantVol} in qubit platforms such as trapped ions~\cite{Zhang2019-uv, Monroe2014-tr}, superconducting circuits~\cite{stein2023microarchitectures,Zhou2022-rw} and color centers in diamond~\cite{Nemoto2016-sf}.  
On the other hand, large distance-scale quantum networks are key to enabling distributed quantum information processing~\cite{Wolk2021-ci} with applications in quantum data center networks~\cite{Liu2023-vj}, secure delegated quantum computation in the cloud~\cite{Fitzsimons2017-zk}, quantum key distribution networks~\cite{Mehic2020-rb}, and quantum sensor networks~\cite{Malia2022-zh,Zhang2021-ow,Guo2019-ed}.

Quantum communication that is needed for networking distinct quantum nodes irrespective of the distance scale is ubiquitously implemented using light. 
The various degrees of freedom of single photons such as the polarization, time-bin, or spatio-spectro-temporal mode provide means to encode quantum information in light~\cite{Slussarenko2019-ih,Flamini2019-nn}. 
Given the relative ease of generating single photon quantum states, these form the primary focus of the present-day quantum networking efforts. 
However, modes of the quantized light field themselves on the whole are quantum objects, also referred to as \textit{qumodes} that can be prepared in a myriad possible multi-photon states and be used to encode quantum information more efficiently than with single photons~\cite{Weedbrook2012-xs}. 
Among such possibilities, the Gottesman-Kitaev-Preskill (GKP) bosonic error correcting code is known to be resilient to photon loss~\cite{Gottesman2001-ks}. 
Qubits based on the GKP encoding have been shown to nearly achieve the quantum communication capacity of Gaussian thermal-loss channels under mean photon number constraint~\cite{Noh2019-po}, which model most common transmission media such as optical fiber and free-space links. 
As a result, GKP codes, though difficult to generate, are widely viewed to be the future of quantum communication. 


Quantum networking with light is enabled by specialized helper nodes---\textit{quantum repeaters}~\cite{Munro2015-pa} and \textit{quantum switches} (also referred to sometimes as quantum routers)~\cite{lee2022quantum,vardoyan2023capacity,Tillman2022-kf}, that consist of quantum optical sources and detectors, quantum memories, and fast optical switches. 
They can forward quantum data reliably in the face of photon loss and thermal noise and do so efficiently at rates above direct transmission~\cite{Pirandola2017-cq}. 
While the former are line elements connecting two \textit{clients} (i.e., nodes that are directly attached to it, which could be end users or other repeaters or switches), the latter have the additional ability to switch between, or connect, or correlate, multiple clients. 
Together, they can be used to realize quantum networks of arbitrary topology at different distance scales.

Quantum repeaters based on the GKP code have been proposed and analyzed recently~\cite{Schmidt2023-dg,RSPJG23,Fukui2021-au,Rozpedek2021-lh}. 
Most notable among these for the entanglement rates it enables is the repeater of Ref.~\cite{RSPJG23}. 
Its architecture involves the use of multiplexed copies of physical-logical GKP qubit resource Bell states. 
The logical qubit part consists of a collection of qubits prepared in the GKP code concatenated with a qubit quantum error correcting code, and is retained at the repeater, serving as an all-photonic quantum memory, while the physical qubit part is simply a single GKP encoded qubit that is transmitted towards a neighboring node for interfacing via physical-physical GKP qubit Bell state measurement (BSM). 
These BSMs generate logical-logical GKP qubit-based \textit{entanglement links} between pairs of adjacent nodes in a repeater chain. 
Ref.~\cite{RSPJG23} considered the logical qubit to be 7 physical GKP qubits encoded in the [[7,1,3]] Steane code. 
In this case, the overall entangled resource state (physical-logical Bell state) is related to an 8-qubit graph state of cube topology. 

On the multiplexed logical-logical entanglement links, the repeaters of Ref.~\cite{RSPJG23} implement an \textit{entanglement ranking-based link matching} protocol, a heuristic where the entanglement links on either side of each repeater are ranked based on the quality of the GKP-qubit entanglement established by the physical-physical GKP BSMs as indicated by the analog outcome values of the BSM, then matched across each repeater node by rank for error-corrected entanglement swapping between the corresponding logical GKP qubits (all-photonic quantum memories). 
A chain of equi-spaced repeaters of this type was shown to support end-to-end entanglement rates as high as 0.7 ebits/mode at total distances as large as 700 km under realistic assumptions for GKP qubit quality expressed in terms of GKP squeezing~\cite{Tzitrin2020-vz} and coherent homodyne detector efficiencies. 

In this paper, we introduce and analyze a quantum switch for GKP-qubit-based entanglement distribution networks, whose architecture is compatible with that of the repeater of Ref.~\cite{RSPJG23}, involving the same multiplexed physical-logical GKP-qubit-based entangled resource states. 
We focus on bipartite entanglement switching, i.e., where the switch facilitates entanglement distribution between pairs of its clients, which may most generally be at different distances. 
We consider a protocol heuristic for the switch that is a multi-client generalization of the entanglement ranking-based link matching protocol heuristic of Ref.~\cite{RSPJG23} where \textit{elementary} entanglement links shared by the switch with the different clients are globally ranked and matched with each other to \textit{connect} clients, or in other words, generate \textit{end-to-end} entanglement links between clients. 
We analyze the protocol for an exemplary data center network, where the data center features as a client of the switch with all the other clients intending to connect to the data center alone and not with each other. 
Given that the entangled resource states form the most valuable commodities at the nodes of the network, we study the problem of optimal allocation of these resources towards the different \textit{connections}, or client pairings, such that the said protocol achieves the maximum sum throughput of the switch (or the total switch rate), while also being fair in terms of the individual entanglement rates under a simple notion of rate fairness that we introduce.

The main contributions of this work can be summarized as follows:
\begin{itemize}
{\item We present an architecture for a GKP-qubit-based quantum switch with multiple clients, where the clients are most generally at different distances from the switch and have different numbers of multiplexed elementary entanglement links with the switch.}

{\item We present a generalization of the entanglement-ranking-based link matching protocol heuristic from Ref.~\cite{RSPJG23} for the switch for bipartite entanglement distribution between clients via entanglement swapping.} 

{\item For the simplest instance of the proposed switch, namely one that connects just one client pair (i.e., enables just one connection) with the two clients being most generally at different distances (essentially an asymmetric repeater), we derive end-to-end entanglement rates based on the so-called 6-state protocol~\cite{DB98}, which is an achievable rate of distilling two-qubit maximally entangled states from the end-to-end entanglement states distributed by the switch.}

{\item For the two-client (single connection) switch, given a total number of GKP-qubit-based entangled resource states per round of the switch protocol, we numerically determine the optimal resource allocation with the connection, i.e., among the two clients that maximizes end-to-end entanglement rate when the entanglement-ranking-based matching protocol heuristic is used. 
Given the total distance between two clients, we also numerically determine the optimal placement of a switch (repeater) node between the two.}

{\item We define an entanglement-rate fairness measure for a multi-client switch that enables multiple connections. 
We use it to determine the optimal allocation of the entangled resource states per round of the switch protocol towards the different connections supported by the switch in a data center network example such that the sum throughput of the switch is maximized while the different connections receive fair individual entanglement rates.}
\end{itemize}

The paper is organized as follows. 
In Sec.~\ref{sec:switch_arch_prot}, we describe our switch model, architecture, and the generalized entanglement-ranking-based link matching (GERM) protocol heuristic for entanglement swapping.
In Sec.~\ref{sec:two-user}, we begin by analyzing the simplest instance of the proposed quantum switch connecting two users most generally at different distances. 
Here, we make key observations on optimally allocating the GKP-qubit-based entangled resource states at the switch, given a fixed number of resources, and placing the switch nodes, given a fixed total distance between users, to achieve the best sum entanglement rates across the repeater. 
In Sec.~\ref{sec:multi-client}, we describe and analyze an illustrative yet comprehensive example of a multi-client quantum switch, namely, a data-center network. 
Here, we introduce our fairness metric and determine the optimal allocation of resource states that yield the highest total switch rate while providing fairness between all the entanglement connections. 
We conclude with a summary and general guidelines for the proposed quantum switch in Sec.~\ref{sec: conclusion}~\footnote{See github.com/mohadesehazari98/Quantum\textunderscore Switch for the Matlab simulations.}.


\section{SWITCH MODEL, ARCHITECTURE AND PROTOCOL}
\label{sec:switch_arch_prot}
\begin{figure}
    \centering
    \includegraphics[width=0.9\columnwidth]{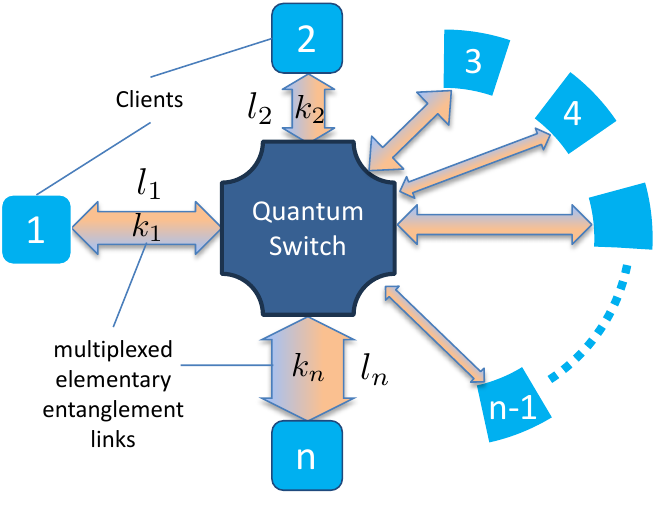}
    \caption{A generic quantum switch with $n$ clients at different distances $l_i$ and with different number $k_i$ of multiplexed elementary entanglement links with the switch, where $i\in\{1,\ldots,n\}$.
    }
\label{fig:schematic}
\end{figure}

\noindent\textbf{MODEL:} Consider the general layout of a quantum switch depicted in Fig.~\ref{fig:schematic}. 
We model the switch as having multiple ($n$) clients that are at different distances ($l_1,\ldots,l_n$), each attempting a different number ($k_1,\ldots,k_n$) of multiplexed elementary entanglement link generation with it periodically at a set global clock rate. 
The role of the switch is to connect these elementary entanglement links to generate end-to-end entanglement links between different pairs of clients. 

\noindent\textbf{ARCHITECTURE:} The architecture of the proposed switch is primarily based on the GKP code. 
The $d$-dimensional GKP code, in short, is a bosonic quantum error correcting code whose codespace is a $d$-dimensional qudit subspace of the infinite-dimensional Hilbert space associated with a single mode of the light field that has intrinsic resilience against photon loss. 
It is defined by a couple of syndrome measurements, which, along with conditional correction operations, can be repeatedly performed to preserve a quantum state in the codespace in the face of photon loss. 
Here, we focus on the $d=2$ code, i.e., GKP qubits. 
For more details on the quantum physical description of realistic GKP qubits, associated noise characteristics (due to finite squeezing effects), and error correction properties, c.f., Ref.~\cite{RSPJG23}.

The architecture involves elementary entanglement link generation and storage across the switch and each of its clients using GKP-qubit-based entangled resource states, which are essentially physical-logical GKP-qubit Bell states. 
It involves each party (the switch and a client) transmitting the physical GKP qubit part of its entangled resource state, referred to as the \textit{outer} leaf qubits, towards one another, which meet halfway along the length of the transmission line. 
They undergo GKP qubit BSM, resulting in a logical-logical GKP qubit entanglement link with the pair of logical GKP qubits, referred to as the \textit{inner leaf} qubits, emulating quantum memories and thus acting as entanglement storage. 
For example, the logical GKP qubit is chosen to comprise the GKP qubit code concatenated with the [[7,1,3]] Steane code. 
The [[7,1,3]] Steane code is a qubit quantum error correcting code that encodes 1 logical qubit in 7 physical qubits and can correct arbitrary errors on up to any one of the physical qubits. 
For more details on quantum error correcting codes, c.f., Ref.~\cite{Devitt2013-fl}. 
In this case, the entangled resource state is equivalent to an 8-qubit graph state of cube topology up to Hadamard gates on 4 of the eight qubits. 

Another key feature of the architecture is multiplexing, where each client-switch pair attempts to generate multiple elementary entanglement links simultaneously using multiplexed entangled resource states. 
Note that the GKP-qubit BSM is inherently deterministic, and thus, all multiplexed attempts succeed, but the quality of each link depends on the continuous real-valued outcomes of the measurements involved in the BSM. 

Finally, for end-to-end entanglement link generation between a pair of clients, the architecture relies on logical-logical error-corrected entanglement swap operations between their inner leaf qubits at the switch to connect the corresponding elementary entanglement links. 
The hitherto described architecture of the switch is entirely compatible with that of the repeaters of Ref.~\cite{RSPJG23}, which we thus refer the reader to for more details. 

\noindent\textbf{PROTOCOL:} Given multiple elementary entanglement links with each client, the proposed switch deploys a generalized multi-client version of the entanglement-ranking-based link matching protocol of Ref.~\cite{RSPJG23} for generating end-to-end entanglement links. 
To briefly describe the algorithm behind the GERM protocol, consider the $k_i$ multiplexed elementary entanglement links generated between the switch and the $i^\textrm{th}$ client in each protocol round. 
Each of these links is of a different quality of entanglement content, quantified by the likelihood of no logical error on the outer leaf qubits involved in the BSM that generated the link, given by:
\begin{align}
P_{\textrm{no-error}} = (1 - P_p(p_0))(1 - P_q(q_0)).
    \label{eq: no-error}
\end{align}
Here, $P_{p}(p_0)$ and $P_{q}(q_0)$ are the likelihoods of incurring an error (incorrect detection of the GKP-qubit state) when measuring in the $p$ and $q$ quadratures and observing real-valued outcomes $p_0$ and $q_0$, respectively. 
The aggregate collection of all the $\sum_{i=1}^{n} k_i$ elementary entanglement links are then globally ranked and sorted based on the value of their $P_{\textrm{no-error}}$ of Eq.~\ref{eq: no-error}, referred to as the outer leaf error and sorted in descending order. 
Pairs of links identified from top to bottom of the rank list, as long as they are not with the same client, and belong to the set of connections being enabled by the switch, are matched and connected by the switch. 
The algorithm proceeds by removing matched links from the rank list to continue matching subsequent pairs until we are possibly left with links with any one client alone. 
Note that for brevity of analysis, we do not carry forward unused elementary entanglement links to subsequent protocol rounds; they are simply discarded. 
The generalized protocol is described below in terms of a pseudo code.
\begin{algorithm}
\caption{GERM}
\begin{algorithmic}[]
\State $n \gets \textrm{number of users}$
\State $k_i \gets  \textrm{number of multiplexed links with\ } i^{\textrm{th}} \ \textrm{user}$
\State $k_{\textrm{total}} \gets \sum_i k_i: \textrm{\ sum total number of multiplexed links}$
\State $\textrm{ErrMat} \gets \textrm{Merge} \left( P_{\textrm{no-error,}i}, i \in \{1,...,n\}\right)$
\\ 
\Comment{Note that: $\textrm{Size}\left(P_{\textrm{no-error,}i}\right)= k_i, \ \textrm{Size}\left(\textrm{ErrMat}\right)=k_\textrm{total}$}
\State $\textrm{ErrMat} \gets \textrm{descending Sort} \left( \textrm{ErrMat}\right)$

\While{$\textrm{ErrMat} \neq 0$}
    \State $i \gets 2$
    \While{TRUE}
        \State $\textrm{ErrMat}\left( 1\right) \textrm{ is  } P_{\textrm{no-error,}a}\left(b\right)$
        \State $\textrm{ErrMat}\left( i\right) \textrm{ is  } P_{\textrm{no-error,}a'}\left(b'\right)$
        \If{($a\leftrightarrow a'$) is a required connection} 
            \State $\bold{Connect}$ user $a$'s $b^{\textrm{th}}$ link to user $a'$'s $b'^{\textrm{th}}$ link
            \State $\bold{Remove}$ $\textrm{ErrMat}\left( 1\right)$ and $\textrm{ErrMat}\left( i\right)$
            \State $\bold{Break}$
        \Else
            \State $i \gets i+1$
        \EndIf 
    \EndWhile
\EndWhile
\end{algorithmic}
\end{algorithm}

\section{A SWITCH CONNECTING TWO CLIENTS}
\label{sec:two-user}

\begin{figure*}
    \centering
    \includegraphics[width=\textwidth]{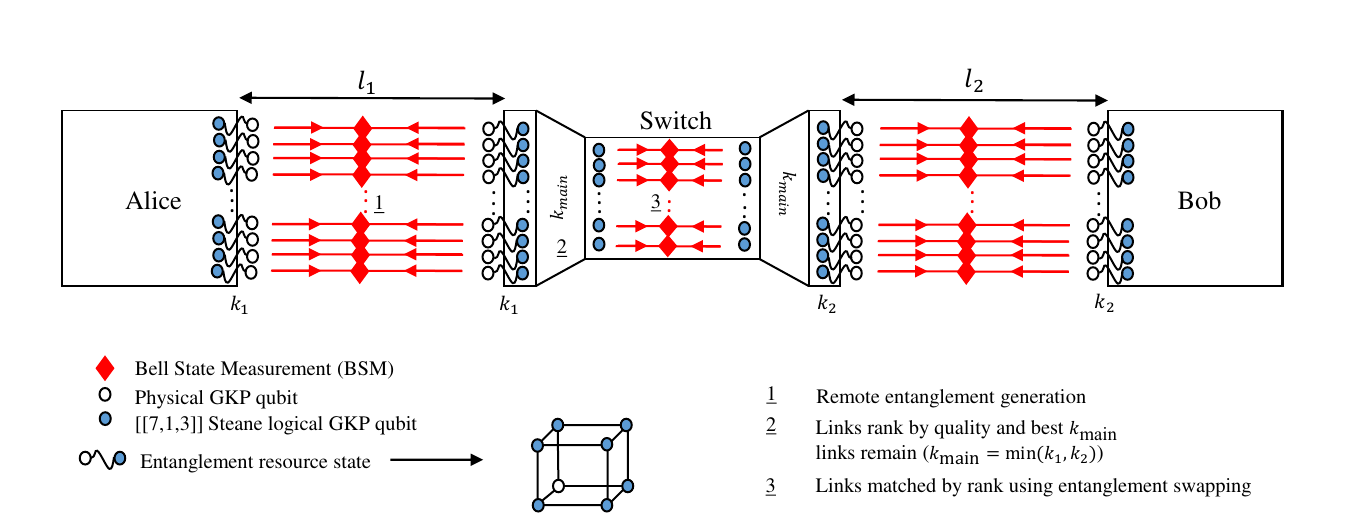}
    \caption{
    The proposed multiplexed all-photonic quantum switch based on GKP-encoded qubits and the [[7,1,3]] Steane code in its simplest form consisting of just two clients. The clients are located at distances $(l_1,l_2)$ from the switch. The switch prepares $k_{\textrm{total}} = k_{1} (\textrm{left}) + k_{2} (\textrm{right})$ entangled resource states that correspond on the logical level to a Bell pair between a concatenated-coded qubit and a bare GKP qubit and on the physical level to a cube graph state of eight GKP-qubits, with the clients matching the respective preparations from their end.  Remote entanglement generation is performed between the switch and each client by sending the bare physical GKP-qubits(white/empty circle) toward each other for Bell State Measurement (BSM). The elementary entanglement links thus generated are ranked according to their reliability estimated from the GKP syndromes obtained from the continuous BSM outcomes. This ranking information, as well as logical BSM outcomes, are sent to the switch. The switch chooses the best $k_{\textrm{main}} = \min(k_{1},k_{2})$ links from each channel (left and right) to perform entanglement swapping on the concatenated-coded qubits(blue/filled circle) based on that ranking information.
    }
    \label{fig:asy_two_cli}
\end{figure*}

Here we analyze the simplest instance of the proposed switch, namely, one with just two clients that are most generally at different distances ($l_1,l_2$) and having different number ($k_1,k_2$) of multiplexed elementary entanglement links with the switch, as depicted in Fig.~\ref{fig:asy_two_cli}. 
We will first 
derive an expression for the end-to-end entanglement rate in terms of ebits, or maximally entangled qubit pairs that can be distilled from the end-to-end entanglement links generated per round of the switch protocol. 
Subsequently, for any given total number of multiplexed elementary entanglement links $k_{\textrm{total}}$ that the switch can create with the two clients per protocol round, we will optimize the end-to-end entanglement rate over all possible allocations such that $k_1+k_2=k_{\textrm{total}}$.
\subsection{END-TO-END ENTANGLEMENT RATES}

First of all, we note that when $k_1 \neq k_2$,  the overall number of end-to-end entanglement links that the switch can facilitate is $k_{\textrm{main}}=\min(k_1, k_2)$, where only the best $k_{\textrm{main}}$ links of the client with a larger number of links get used.
Secondly, in generating the multiplexed entanglement links between the switch and the two clients, the outer leaf qubits travel $l_1/2$ and $l_2/2$, respectively. 
Meanwhile, the corresponding inner leaf qubits travel in local fiber spools twice the distance as the outer leaf qubits do, i.e., $l_1$ and $l_2$, respectively. 
This is because they serve as quantum memory until the information about the quality of the generated links (the analog outcomes of the BSM) reaches the switch. 
When $l_1 \neq l_2$, to ensure that the switch has the ranking information of its entanglement links with both the clients, the inner leaf qubits of elementary entanglement links associated with both clients travel an optical fiber of length given by $\max(l_1, l_2)$. 

Thirdly, the quality of the entangled resource states used for generating the entanglement links with the two clients in terms of their GKP squeezing and the associated logical error probabilities need be no better than the worst of the probabilities of logical error caused on the inner leaf qubits by the optical fiber transmission distances $l_1$ and $l_2$, respectively. 
This is because the error probabilities add up anyway. 
The larger the distance, the greater the associated error probability. 
The quality of the entangled resource state on the other hand is governed by the size of the discard window $\nu$ used in GKP qubit measurements in its preparation (more discussed in Appendix~\ref{sec:App_1}). 
The smaller the discard window the larger the logical error probabilities in the entangled resource state qubits.

In any given protocol round, consider the $j^\textrm{th}$ end-to-end entanglement link, which is created by entanglement swapping the $j^\textrm{th}$-ranked elementary entanglement links with the two clients at the switch, where $j\in (1,\ldots,k_{\textrm{main}})$. 
The rate of generating perfect ebits from this end-to-end entanglement link is a function of the inner and outer leaf logical error probabilities $Q_{X/Z, \textrm{outer},(i)} (j), Q_{X/Z, \textrm{inner},(i)} (j)$ of the parent  elementary entanglement links associated with each of the two clients $i\in \{1,2\}$. 
Here $Q_{X/Z, \text{outer}, (i)}$ is a $k_i$-dimensional vector which holds the logical error probability on the outer leaf qubits of the elementary entanglement link between the $i^{\textrm{th}}$ client and the switch. 
On the other hand, $Q_{X/Z, \text{inner}, (i)}$ is a $k_i \times 2$ matrix which holds the logical error probability of the corresponding inner leaf qubits. 
Given that the inner leaf GKP qubits are further error-protected by a higher lever error correcting code (ECC), namely, the Steane code, the end users have the ability to also use the error syndromes of the Steane code (along with those of the GKP qubit syndromes) to find the best end-to-end links. 
The first column represents the inner leaf logical error when there is no Steane code error syndrome ($s=0$) detected, while the second represents the case where there is a Steane code error syndrome ($s=1$). 
The total error probability of the $j^\textrm{th}$-ranked elementary entanglement link with the $i^{\textrm{th}}$ client with Steane-code-level error syndrome $s$ is given by
\begin{equation}
\begin{aligned}
    Q_{X/Z, (i)}&(s,j) = Q_{X/Z, \text{inner}, (i)}(s)\left(1-Q_{X/Z, \text{outer}, (i)}(j)\right) \\
    &+ \left(1- Q_{X/Z, \text{inner}, (i)}(s)\right)Q_{X/Z, \text{outer}, (i)}(j).
    \end{aligned}
\end{equation}

To classify end-to-end entanglement links based on the Steane-code-level error syndromes ($s=1$) of their parent elementary entanglement links, we define vectors $\vec{m}_{X/Z}=(\{m_{X/Z}(i):i\in\{1,2\}\})$. 
We note that
\begin{equation}
    ||\Vec{m}_{X/Z}||_1 = c_{X/Z}\in \{0,1,2\}.
\label{eq:m_vector}
\end{equation}
The total error probability associated with the  $j^{th}$-ranked end-to-end entanglement link and a given $\vec{m}_{X/Z}$ is then given by
\begin{align}
    Q_{X/Z,\textrm{end}}&(\Vec{m}_{X/Z},j) = (1/2) \left(1- \prod_{i=1}^2 \right. \nonumber \\
    &\left(1-2Q_{X/Z,(i)}(s=1,j)\right)^{m_{X/Z}(i)}\nonumber \\
    &\left. \left(1-2Q_{X/Z,(i)}(s=0,j)\right)^{1-m_{X/Z}(i)} \right).
    \label{eq:elementaryLinkFlip}
\end{align}
The total end-to-end entanglement rate of $k_{\textrm{main}}$ multiplexed links when distilled separately from the different $j$'s and the different $\Vec{m}_{X/Z}$'s is given by:
\begin{align} &R_{\textrm{e2e}} = \sum_{j=1}^{k_{\textrm{main}}}\sum_{\vec{m}_X,\vec{m}_Z} p_{X}(\vec{m}_X)p_{Z}(\vec{m}_Z)\nonumber\\ &r\left(Q_{X,\textrm{end}}(\Vec{m}_X,j), Q_{Z,\textrm{end}}(\vec{m}_Z,j)\right)\label{eq:re2e}.
\end{align}
Here $r$ is the secret-key fraction or a lower bound on distillable entanglement. 
The quantity $p_{X/Z}(\vec{m}_{X/Z})$ is the probability of the parent elementary entanglement links having error syndromes $\vec{m}_{X/Z}$ on their inner leaves, given by 
\begin{align}
    p_{X/Z}(\Vec{m}_{X/Z}) =\prod_{i=1}^2 t_{X/Z, (i)}^{m_{X/Z}(i)}\left(1 - t_{X/Z, (i)}\right)^{1-m_{X/Z}(i)},
      \label{eq: t}
    \end{align}
where $t_{X/Z, (i)}$ is the probability of an error syndrome ($s=1$) on the inner leaves of the elementary link with the $i^\textrm{th}$ client. 
The quantities $t_{X/Z, (i)}$, $Q_{X/Z, \textrm{inner}, (i)}(s=\{0,1\})$ and $Q_{X/Z, \text{outer}, (i)}(j=\{1,...,k_{i}\})$ are all obtained through simulation. 

Note that the above derivation can be easy generalized to a chain of quantum repeaters with asymmetric spacings.

\subsection{OPTIMIZING RESOURCE ALLOCATION}
\begin{figure}
\centering
\subfloat[\label{1a}]{%
    \includegraphics[width=7.5cm, height=6cm]{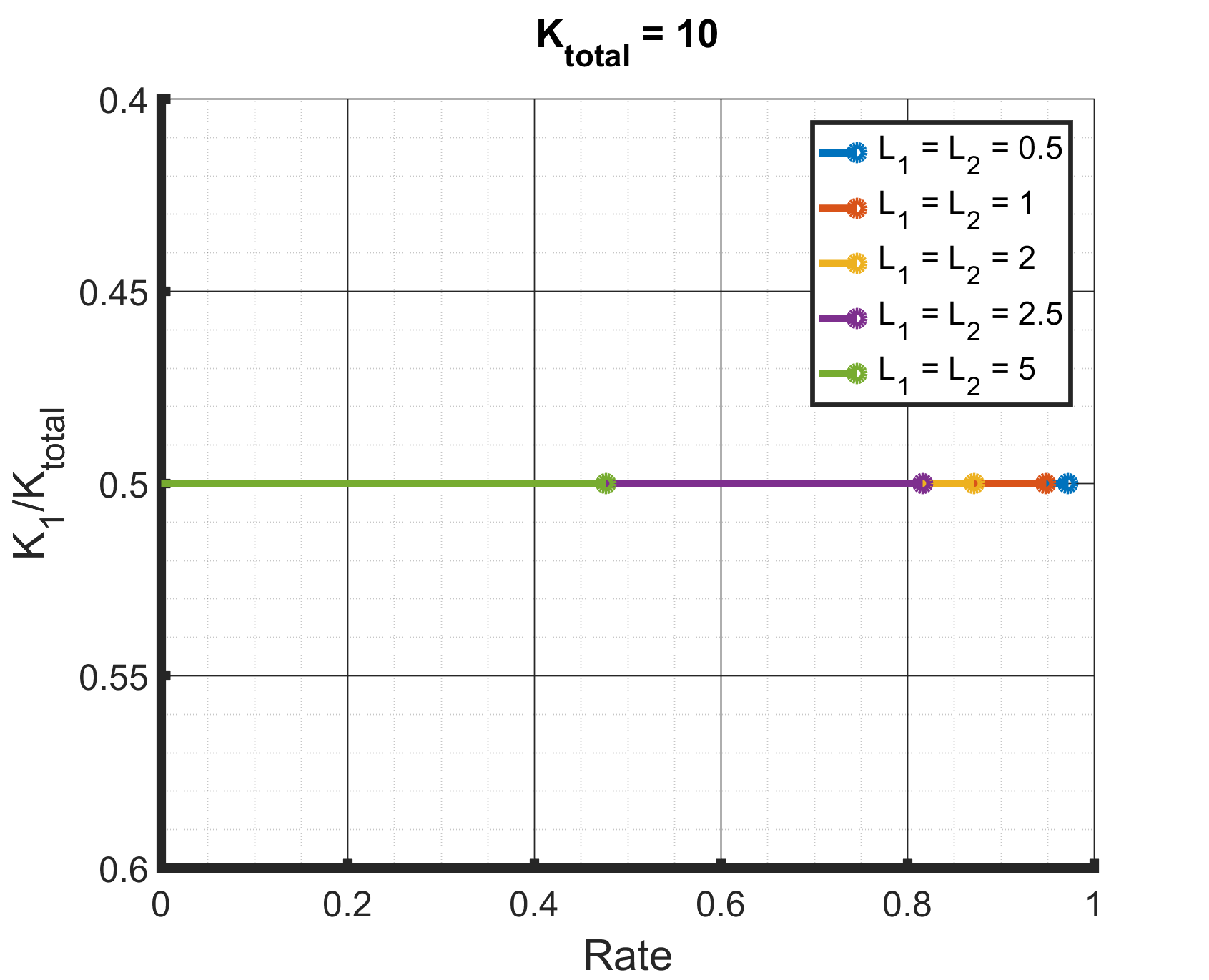}}
    \\
\subfloat[\label{1b}]{%
    \includegraphics[width=7.5cm, height=6cm]{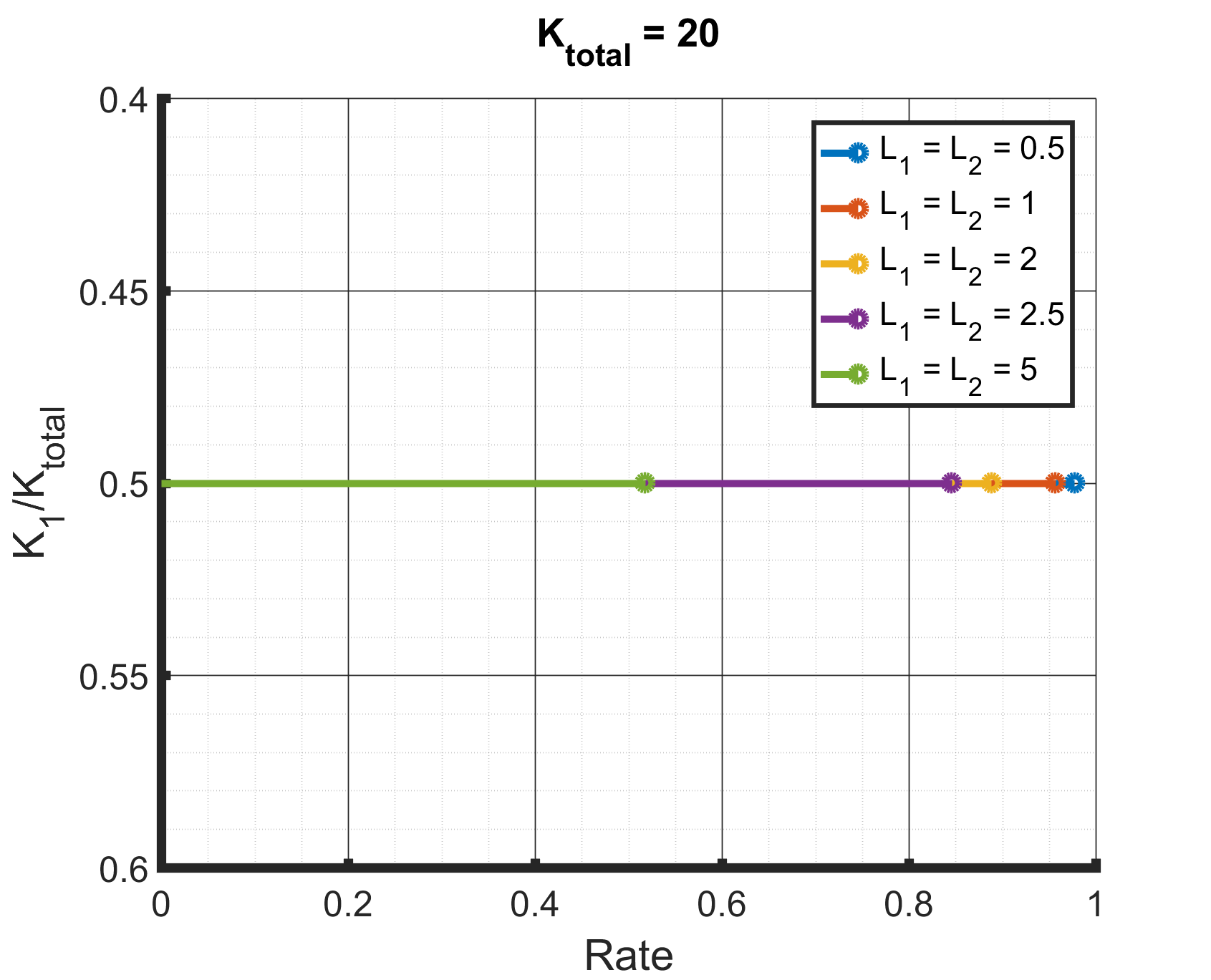}}
    \\
\subfloat[\label{1b}]{%
    \includegraphics[width=7.5cm, height=6cm]{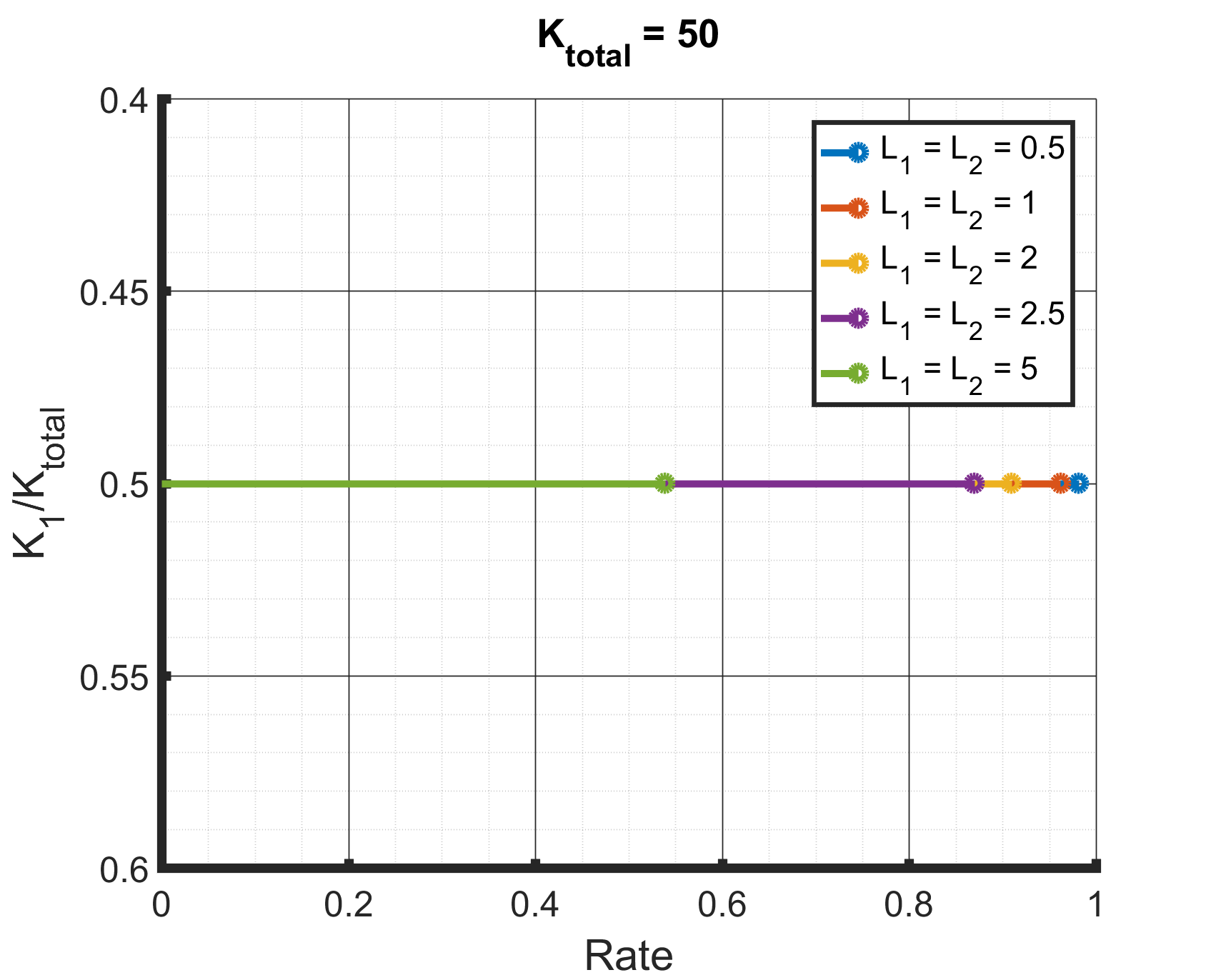}}    
\caption{
Rate saturation with the increase in total number of resource states while optimally shared between two clients, i.e., $l_1 = l_2 = {l_{\textrm{total}}/2}$ ($l_i \in \{0.5, 1, 2, 2.5, 5\} \textrm{km}$). For every setting ($l_\textrm{total}, k_{\textrm{total}}$), the maximum rate per mode $2R_{\textrm{e2e}}/ k_{\textrm{total}}$ corresponding to the optimum setting ($k_1, k_2$) is plotted on the X-axis. The optimum allocation, regardless of the distance, is found to be $k_1 = k_2 = {k_{\textrm{total}}/2}$, i.e., ${k_1/k_{\textrm{total}}}=0.5$.
}
\label{fig:two_client_simple}
\end{figure}

\begin{figure}
\centering
\subfloat[\label{1a}]{%
    \includegraphics[width=7.5cm, height=6cm]{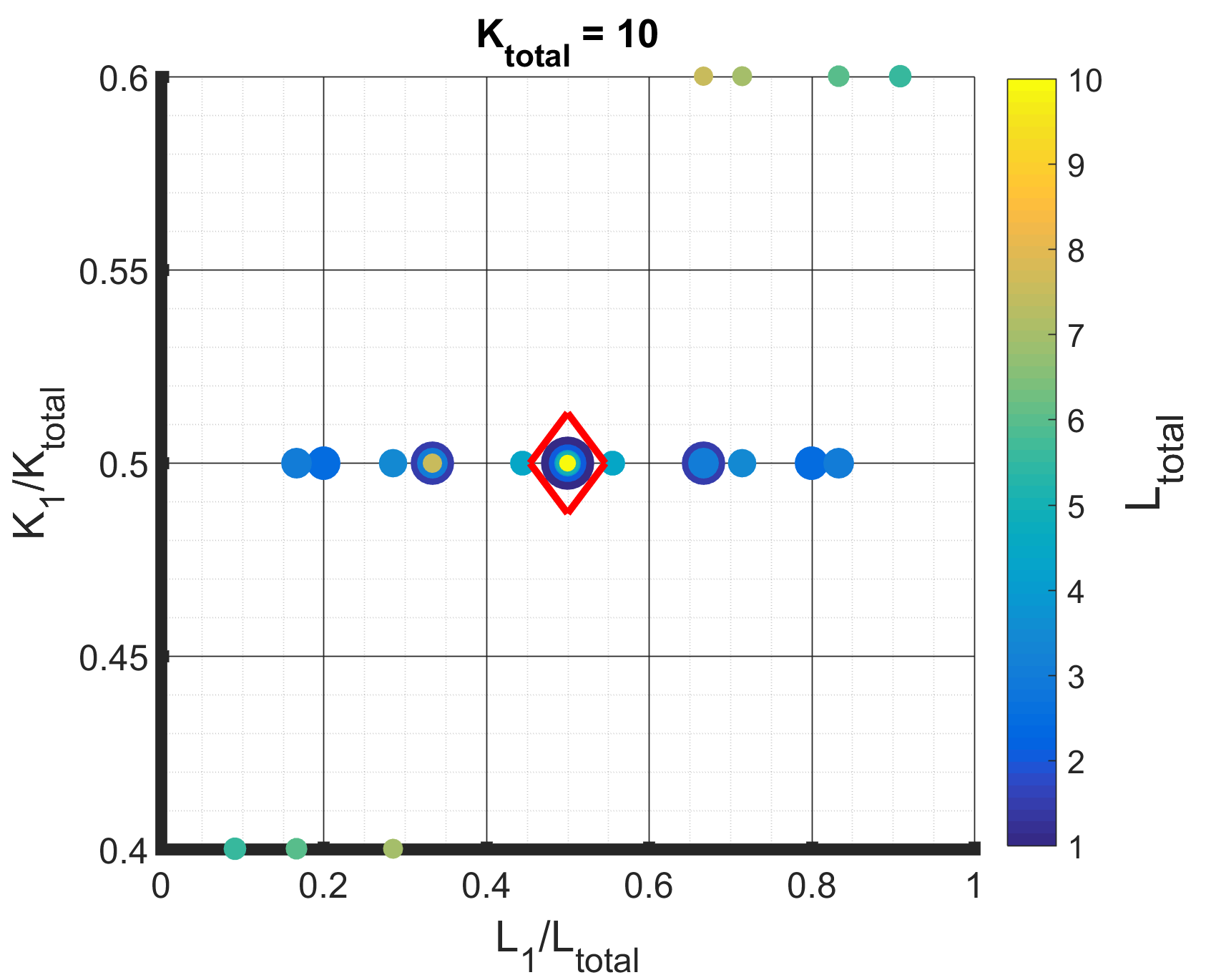}}
    \\
\subfloat[\label{1b}]{%
    \includegraphics[width=7.5cm, height=6cm]{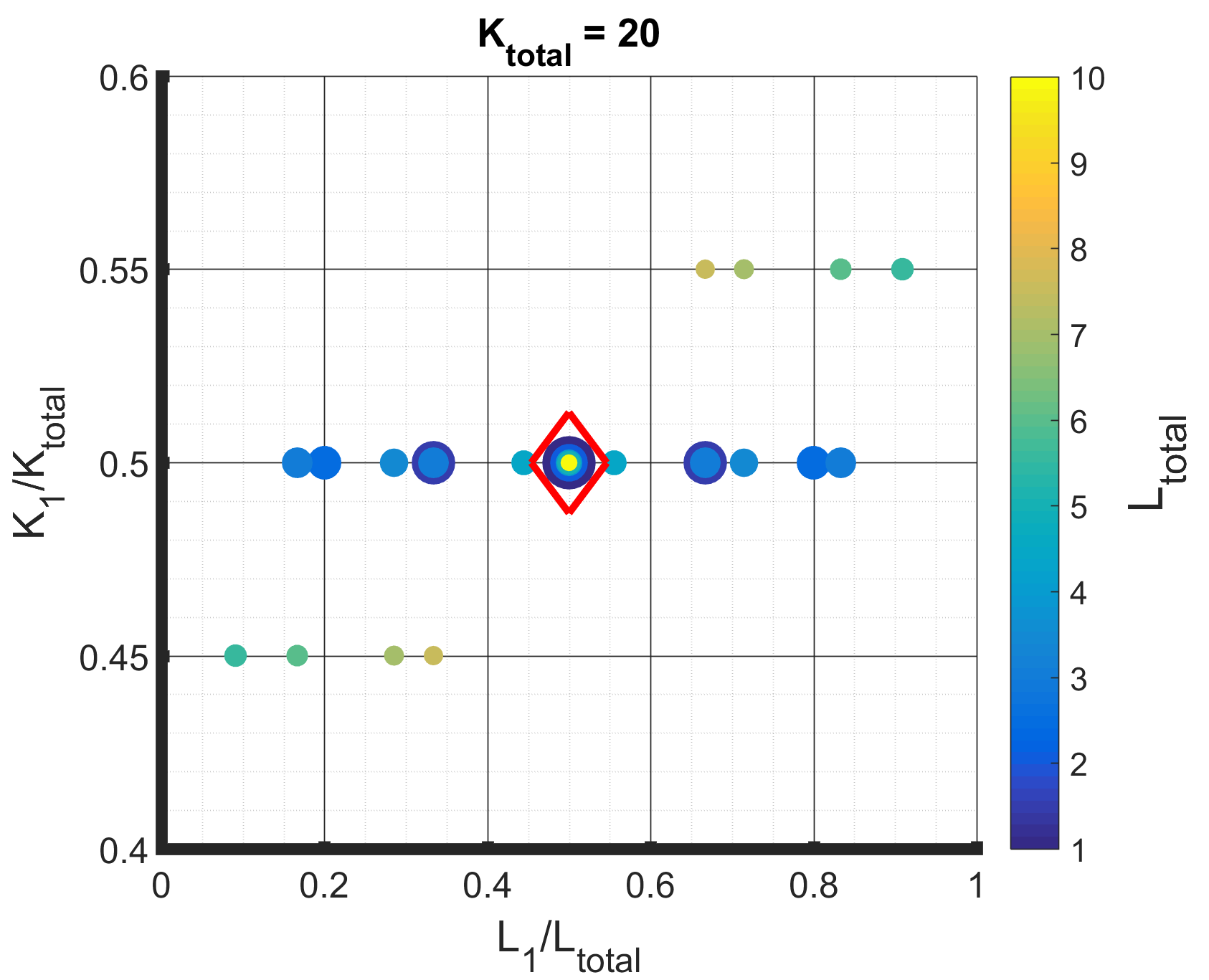}}
    \\
\subfloat[\label{1b}]{%
    \includegraphics[width=7.5cm, height=6cm]{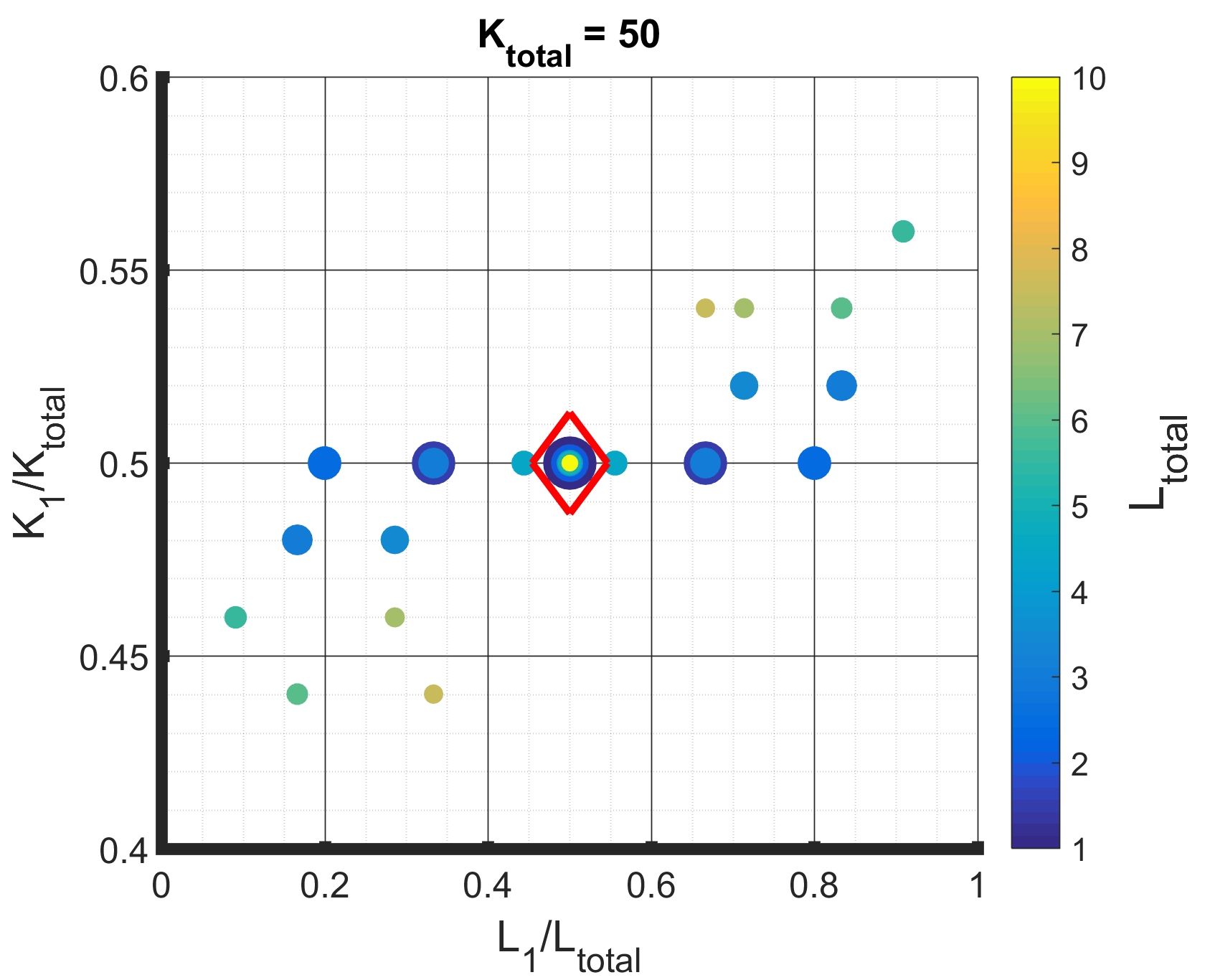}}
\caption{The general case of a two-client switch ($\textrm{client}$-1, $\textrm{client}$-2), where $l_1 + l_2 = l_{\textrm{total}}$ ($l_i \in \{0.5, 1, 2, 2.5, 5\} \textrm{km}$). 
For every setting ($l_1, l_2, k_{\textrm{total}}$), the optimum resource allocation $(k_1,k_2)$, tracked by the number of entangled resource states assigned to $\textrm{client}$-1 over the total number of entangled resource states, i.e., $k_1/k_{\textrm{total}}$ is plotted on the Y-axis, while the distance of $\textrm{client}$-1 over the total distance of the setting ($l_1/l_{\textrm{total}}$) is plotted on the X-axis. The optimum allocation is found to be $k_1 = k_2 = {k_{\textrm{total}}}/2$, i.e., ${k_1/k_{\textrm{total}}}=0.5$ with small deviations at large total distances. 
The red diamond also shows that the maximum total rate $R_{\textrm{e2e}}$ belongs to $l_1=l_2={l_{\textrm{total}}/2}, k_1=k_2={k_{\textrm{total}}/2}$. }
\label{fig:two_client_general}
\end{figure}

For any given resource budget (at the switch), we now look to identify optimal allocation that maximizes the end-to-end entanglement rate of the proposed switch per protocol round. 
In other words, given two clients at distances $(l_1,l_2)$ and a total number $k_{\textrm{total}}$ of elementary entanglement links that the switch can create cumulatively with the clients in a protocol round, we are interested in:
\begin{align}
\max_{k_1, k_2} \ &{R_{\textrm{e2e}} (k_1, k_2;l_1, l_2)} &\nonumber
\\
\textrm{ s.t. } &k_1 + k_2 = k_{\textrm{total}},&\nonumber \\
&k_1,k_2\in \mathbb{Z}^+.&
    \label{eq: solve_k}
\end{align}

\subsection{RESULTS}
We numerically simulated the performance to evaluate the total end-to-end entanglement rate (in terms of ebits) of the proposed all-optical multiplexed quantum switch for two clients at distances $(l_1, l_2)$ where $l_i \in \{0.5, 1, 2, 2.5, 5\}\textrm{km} \forall i\ \in\{1,2\}$ and for different values of total number of entangled resource states $k_{\textrm{total}} \in \{10, 20, 50\}$.
Figure.~\ref{fig:two_client_simple} shows the symmetric case of clients having the same distance from the switch ($l_1=l_2=l$) for different values of $l$ and $k_{\textrm{total}}$, and plots the optimal allocation $(k_1,k_2)$ (tracked by the ratio $k_1/k_{\textrm{total}}$) that maximizes the end-to-end entanglement rate (tracked by entanglement rate per mode $2R_{\textrm{e2e}}/k_{\textrm{total}}$). 
As is expected, equal allocation $k_1/k_{\textrm{total}}=0.5$ is found to be optimal. 
Also, comparing the results over different $k_{\textrm{total}}$ reveals that the end-to-end rate increases as we increase the total resource budget, but then quickly saturates. 
Figure~\ref{fig:two_client_general} shows the resource allocations that maximize the total end-to-end rate for more general settings beyond $l_1=l_2$. 
The size of the circles in the figure is a function of the total distance $l_1 + l_2$. 
As you can see with small deviations, the maximum rate belongs to symmetric allocation, i.e, $k_1 = k_2 = k_{\textrm{total}}/2$.
These deviations, mostly in larger distances, are close enough to be considered as a symmetric allocation ${(k_1 - k_2)}/{(2 k_{\textrm{total}})} \ll \epsilon_{k_{\textrm{total}}}$ with $\epsilon_{10} = 0.1$ and $\epsilon_{50} = 0.06$. 
Not only does the symmetric allocation of resources turn out to be optimal, but also the symmetric disposition of clients about the switch for any fixed $l_{\textrm{total}}$ yields the highest rate. 
To keep the figure simple, we don't show the third dimension, which is $R_{\textrm{e2e}}$. However, the optimal value of the latter quantity is marked as belonging to $\{l_1=l_2, k_1=k_2\}$.
So to operate the proposed quantum switch in its simplest instance with just two clients and with $k_{\textrm{total}}$ entangled resource states, the setting that maximizes the end-to-end rate is to place the repeater in the middle, i.e. $l_1=l_2=l_\textrm{total}/2$ and allocate resources equally, i.e.,  $k_1=k_2=k_\textrm{total}/2$.

\begin{figure*}
    \centering
    \includegraphics[width=0.91\textwidth]{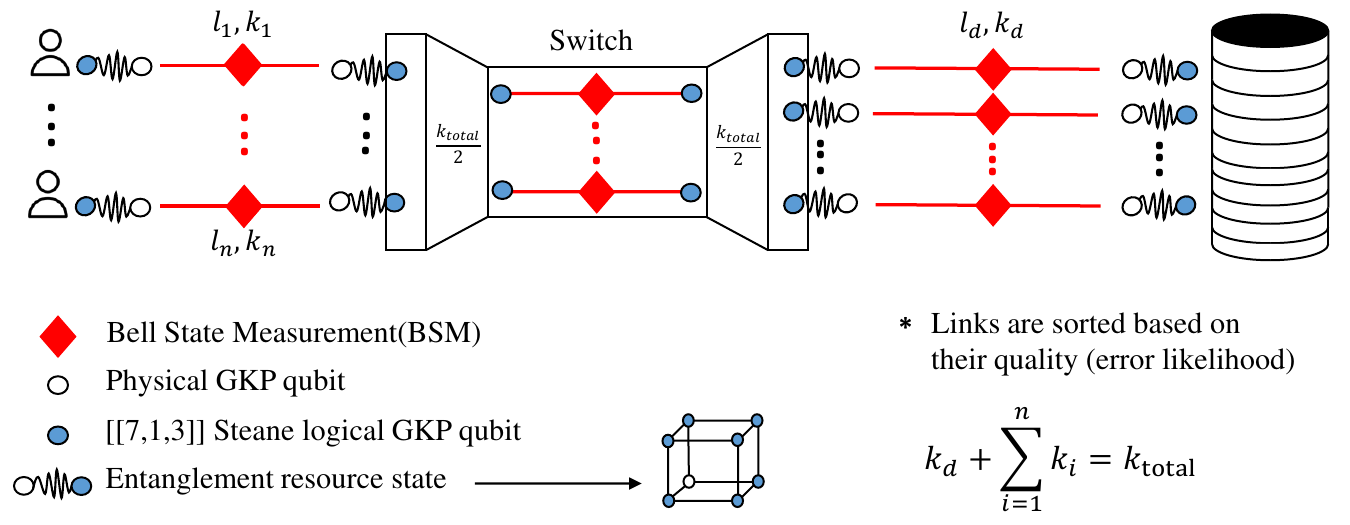}
    \caption{
    Architecture of the proposed multiplexed all-photonic quantum switch in a multi-client data-center network scenario. The data center-switch elementary entanglement links are ranked separately from the elementary links between the switch and the rest of the clients put together. Links from these two rank lists are matched and connected via entanglement swapping at the switch to generate end-to-end entanglement links.  
    }
    \label{fig:asy_multi_cli}
\end{figure*} 

Let's study a case where $\textrm{client}$-1 has a distance of $l_1 = 0.5\textrm{km}$ whereas $\textrm{client}$-2 has a larger distance of $l_2 = 5\textrm{km}$. 
The effect of increasing the number client-2's multiplexed links is that the error likelihood of its best elementary entanglement links would decrease, meaning $\textrm{client}$-2 will have more ``good'' links in hand than before. 
Since $\textrm{client}$-1 is already close to the switch, the error likelihood is guaranteed to be small. 
On the other hand, when $k_1 \neq k_2$, the switch will have to throw away precious $\max(k_1,k_2) - \min(k_1,k_2)$ number of entangled resource states. 
Our simulation results show that the symmetric allocation $k_1=k_2$ will result in a higher total end-to-end rate than the asymmetric allocation $k_1<k_2$. 
In other words, the downside of increasing $k_2$ far exceeds its benefits, because, as mentioned earlier, resource state generation is the most expensive part of a GKP switch architecture. 
Thus, employing all the entangled resource states, i.e. $k_1=k_2$, yields the best total end-to-end entanglement rate for the switch. 


\section{A SWITCH CONNECTING MULTIPLE CLIENTS}
\label{sec:multi-client}

Having described the simplest two-client (i.e., single connection) switch and determined the optimal allocation of resources for maximum end-to-end entanglement rate, we now move to multiple clients. 
To best elucidate the workings of the proposed switch in a multi-client scenario, we will focus on a particular example, namely, that of a data-center switch network where one of the clients of the switch is a \textit{data center} and all the other clients look to connect to the data center alone, as depicted in Fig.~\ref{fig:asy_multi_cli}. 
We will first describe the data center switch network setting along with some nuances of switch operation in the setting and our performance metrics. 
The latter includes the sum throughput of switch and a measure of entanglement rate fairness between the different clients, which we define. 
Given a total entangled resource state budget at the switch, we will then optimize the sum throughput of the switch over the set of all possible resource allocations under an entanglement fairness constraint. 

\subsection{DATA CENTER SWITCH NETWORK: OPERATIONS AND PERFORMANCE METRICS}
Consider a switch network with a data center and $n$ other clients, at distances $(l_d,l_1,\ldots,l_n)$ from the switch as shown in Fig.~\ref{fig:asy_multi_cli}. 
Let the number of elementary links between the switch and the data center be $k_d$, and between the switch and the other clients be $k_i\ \forall i\in\{1,\ldots,n\}$ such that $k_d+\sum_{i=1}^n k_i=k_\textrm{total}$, where $k_\textrm{total}$ is the total resource budget at the switch. 
Given that all connections enabled by the switch here feature a common client, namely, the data center, the GERM protocol (described in Sec.~\ref{sec:switch_arch_prot}) in this case reduces to a special form, where the links that the switch shares with the data center can be ranked separately from all the rest of the links it shares with the other clients. 
The switch can then match the two ranked lists, namely, those of the data center links and those of all the other clients put together, to generate end-to-end entanglement links.
Note that the total number of end-to-end entanglement links that can be generated is given by $\min\{k_d, \sum_{i=1}^n k_i\}$. 

We note that in operating the multi-client switch, to ensure that the switch receives all the outer leaf ranking information required to implement the ranking-based matching protocol, all the inner leaf qubits at the switch should be stored for a time commensurate to light travelling a local fiber spool of length $\max\{l_1,l_2,...,l_n,l_d\}$. 
Also, regarding the quality of the GKP qubits in terms of GKP squeezing that is required in the preparation of the resource states at the first place, as already explained in Section~\ref{sec:two-user}, the logical error probabilities due to the finite GKP squeezing in the individual GKP qubits that are governed by the discard window size $\nu$ need be no better that the logical error associated with the longest optical fiber transmission of the inner leaf qubits during local all-optical storage, namely, $\max\{l_1,l_2,...,l_n,l_d\}$.

The figure of merit in the multi-client scenario, namely, the sum throughput of the switch, is simply the sum of rates of all the individual end-to-end entanglement links that the switch enables, which in the case of the data-center network example, is given by
\begin{align}
R_{\textrm{s}}=\sum_{i=1}^n R_{\textrm{e2e}}^{(i)} (k_d,k_i;l_d,l_i).\label{rsum}
\end{align}
Here, the end-to-end entanglement rate of each individual connection $R_{\textrm{e2e}}^{(i)}\forall i \in\{1,\ldots,n\}$ is defined and evaluated just as in Eq.~\ref{eq:re2e}. 



We further define a fairness measure $F$ that captures how close the end-to-end entanglement rates of the different connections enabled by the switch are to one another. 
The measure $F$ is defined taking the Euclidean distance between the different clients' end-to-end rates ($R_i\equiv R_{e2e}^{(i)}$) scaled by their average:
\begin{align}
    &F(R_1,\ldots,R_n) = \frac{d(R_1,R_2,...,R_n)}{\langle R_1,R_2,...,R_n \rangle},
    \label{eq: Fairness}\\
     &d^2(R_1,R_2,...,R_n) = {\frac{1}{2}\sum_{{i,j}=1}^{n} {|R_i - R_j|}^2}.
    \label{eq: fair_function}
\end{align}
The smaller the fairness measure $F$, the closer the rates among clients are to each other (note that $0 \leq d^2(R_1,...,R_n) \leq n$).
The motivation behind this metric is to ensure that while optimizing the sum throughput, none of the farther clients at a disadvantage compared to nearer clients concerning end-to-end entanglement rates with the data center are left under-served or, worse, totally unserved by the switch.

\subsection{OPTIMIZING RESOURCE ALLOCATION}
\begin{figure*}
    \centering
    \includegraphics[width=1\textwidth]{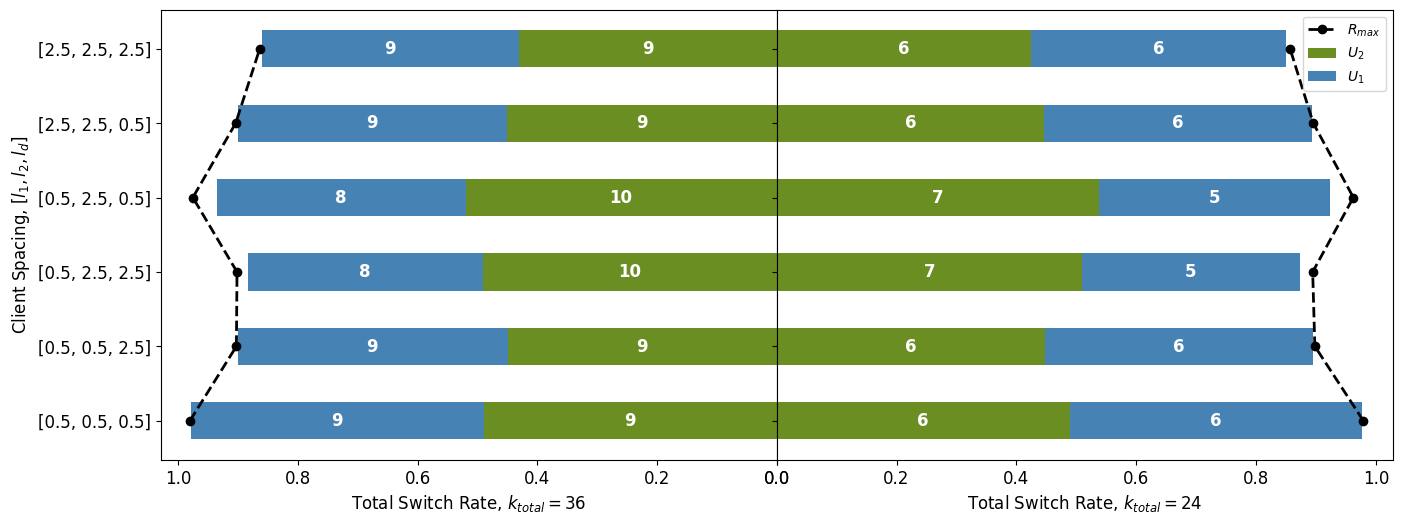}
    \caption{
    The rate analysis of a two-client, one-data center network over various user spacing [$l_1,l_2,l_3$]. The overall switch rate of the fair resource allocation is presented as the bar's height along the x-axis for two different values of $k_{\textrm{total}}$. The maximum overall switch rate is the black dashed line above the bars. ($\frac{k_1}{k_1+k_2}$)/($\frac{k_2}{k_1+k_2}$) of each bar is colored with blue/green to show the resource distribution between the two clients.  
    }
    \label{fig: fair allocation}
\end{figure*}

Similarly to Sec.~\ref{sec:two-user}, for a data-center switch network with the data center and the other clients at distances $(l_d,l_1,\ldots,l_n)$ and any given resource budget $k_{\textrm{total}}$ at the switch, we now look to identify optimal allocation $(k_d,k_1,\ldots,k_n)$ that maximizes the sum throughput $R_{\textrm{s}}$ of Eq.~\ref{rsum} of the proposed switch. 
We do so under the constraint that the fairness measure $F$ of the set of individual client entanglement rates is bounded by a threshold $F_{\textrm{t}}$, whose value is chosen suitably. 
In other words, we are interested in:

\begin{align}
\max_{k_d, k_1,\ldots,k_n} \  &{R_{\textrm{s}} (k_d, k_1,\ldots,k_n;l_d, l_1,\ldots,l_n)}, &\nonumber \\
\textrm{s.t. } & k_d + \sum_{i=1}^n k_i = k_{\textrm{total}}, &\nonumber \\
&F(R_1,\ldots,R_n)< F_{\textrm{t}}, &\nonumber \\[4pt]
&[k_d, k_1,\ldots,k_n] \in \mathbb{Z}^+.&
  \label{eq: solve_k_multi}
\end{align}

First of all, recall the observation from Sec.~\ref{sec:two-user}, namely, that in any single connection between clients $A$ and $B$ through the proposed switch with a total resource state budget of $k_{\textrm{total}}$, regardless of ($l_A, l_B$), the setting that maximizes the rate $R_{\textrm{e2e}}$, is $k_A=k_B=k_{\textrm{total}}/2$.  
Based on this, here we can conclude that for optimal operation $k_d = k_{\textrm{total}}/2$. 
This is because, the above observation implies $k_d=\sum_{i=1}^n k_i$, which along with $k_d+\sum_{i=1}^n k_i=k_{\textrm{total}}$ implies the conclusion. 
Therefore, the optimization is really about allocating the remaining half of the resources at the switch towards the other clients (apart from the data center), namely, $k_i, i \in \{1,2,...,n\}$. 
Equivalently, the resource allocation can be thought of as allocating $2k_i\forall i\in\{1,\ldots,n\}$ resources towards each ``client-$i$-data-center connection'',  and dividing them equally between client-$i$ and the data center. 
This will ensure that the GERM protocol when implemented at the switch will leave no links unutilized. 
Moreover, along with the constraint imposed by the fairness measure, it will also help ensure that all connections are duly served by the switch even while trying to maximize the total switch rate. 

\subsection{RESULTS}
\label{sec: result multi}

We performed numerical simulations to identify optimal resource allocations for data center switch networks of different settings. Figure.~\ref{fig: fair allocation} summarizes the analysis of a switch network of two end-user clients (client-1 and client-2) and a data center (client-3), where client-1 and client-2 request to connect to the data center. As the previous section has indicated, equal resource allocation across a single connection leads to the maximum end-to-end rate. This approach could lead to a more efficient use of resources and a better overall experience for all users involved. Accordingly, in a scenario where clients are symmetric, and their error loads are equal ($l_1 = l_2$), the fair distribution of resources, such that $k_1 = k_2$, is anticipated and verified to result in the highest possible switch rate overall. This finding suggests that the fair distribution of available resources is crucial in achieving optimal performance in a symmetric setting.
According to Ref.\cite{RSPJG23}, increasing the number of multiplexed links in a connection will eventually saturate the total end-to-end rate. While adding more multiplexed links to a low error load connection can considerably increase the overall rate, it saturates the rate, leaving no evident effect in a high error load connection. Therefore, when clients are asymmetrically distributed along the switch ($l_1 \neq l_2$), low error load connections should have more resources if the maximum total switch rate is the target. Increasing the number of multiplexed links can help address the poor per-mode rate of an end-to-end connection, thereby achieving desirable fairness.

\begin{figure*}
    \centering
    \includegraphics[width=1\textwidth]{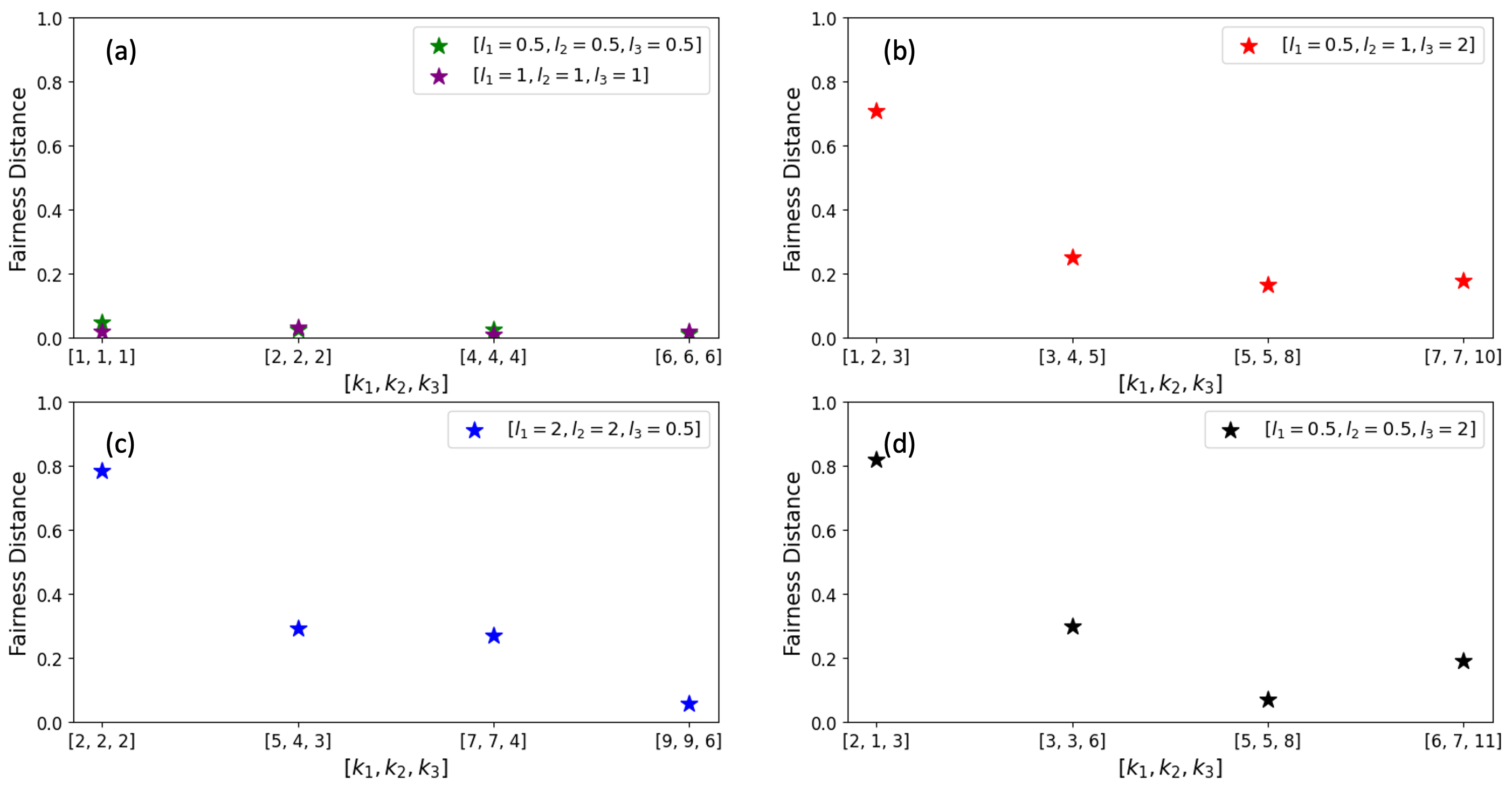}
    \caption{
    The best resource allocation adjusted by the switch to ensure the maximum fairness achievable for the given setting [$l_1, l_2, l_3$] and $k_{\textrm{total}}$, represented on the x-axis along with the corresponding fairness distance on the y-axis. [$l_1, l_2, l_3$] are client distances with data center distance being $l_4=5$km. With a sufficient $k_{\textrm{total}}$, we can increase the fairness by assigning more links to the distant client. However, a closer client always stands high on the ranking and improves its end-to-end rate by performing entanglement swapping with the data center's best links.
    }
    \label{fig: fairness_vs_k}
\end{figure*}

To shift the focus towards the end users, we next study a dominant-data-center example (Fig.~\ref{fig: fairness_vs_k}), where the total switch rate is independent of the client's resource allocation. 
Here we only analyzed the effects of spacing between the end users and the switch, but the analysis can be easily extended to any number of repeaters between the switch and end users.

The rate fairness generally improves as the total number of resource states ($k_{\textrm{total}}$) increases, with farther clients consuming more resource states to maintain end-to-end rates close to other nearer clients. 
The end-to-end rates are more fairly distributed when clients have comparable distances from the switch. 
For instance, a similar resource allocation in [$l_1,l_2,l_3$]=[0.5,1,2]km resulted in better fairness than the [0.5,0.5,2]km setting. 
The fair resource allocation of the symmetric settings like [0.5,0.5,0.5]km or [1,1,1]km develops an equal end-to-end rate for all the clients. 

The dependence of $R_{\textrm{e2e}}$ and in turn of $R_{\textrm{s}}$ on resource allocations $k_i$ turns out to be highly subjective and depends on, e.g., the relative values of the distances between the switch and the clients $l_i$. 
For example, appendix.~\ref{sec: App_2} describes an interesting special case where any one of the clients of the switch is a dominant client in that it is farther from the switch and, as a result, dictates the order of magnitude of the total logical error probabilities, and thereby how the total switch rate $R_{\textrm{s}}$ becomes independent of the other clients' resource budget allocations. 

\section{CONCLUSIONS AND OUTLOOK}
\label{sec: conclusion}

We presented an architecture for a GKP-qubit-based multiplexed all-optical quantum switch for entanglement distribution networks. 
We analyzed the optimal allocation of resources among any number of clients and the optimum relative location of the switch for a switch connecting two clients. 
The optimum was studied separately under the overall switch rate and the fairness of users' overall end-to-end entanglement link generation rates.  
We zoomed in on the network and focused on distances of the inter-nodal spacing orders so that the result of the paper would be useful for any network architecture of end users, repeaters, and switches. 
Based on our findings, here is a summary of guidelines to optimally locate and allocate resources of the quantum network helper nodes, i.e., repeaters and switches, to enhance the overall switch rate and also the fairness of rates among clients:

I. To maximize the generation rate of end-to-end entangled links between two clients connected via a single switch, the relative location of the switch should be in the middle of the connection, and the resources should be shared equally.
When it comes to an end-to-end connection, symmetric resource allocation always results in a better end-to-end rate, even when clients are at different distances from the switch. This applies to single switch connections and chains of switches (or repeaters) that connect two end users.

II. The GERM protocol, short for the generalized entanglement-ranking-based link matching, has been introduced in this work to optimize the entanglement generation throughput across multiple clients and connections. The protocol ensures the highest overall rate by favoring higher-quality connections. It achieves this by minimizing the logical error over its top-quality connections. Implementing this protocol can markedly improve the efficiency of entanglement generation, especially in critical settings where high-speed data transfer is essential.

III. To ensure fairness among clients, the switch should prioritize lower-quality connections by giving them more entangled links to account for their relatively low per-mode rate. This approach will help to ensure that each client's end-to-end rate is equitable, regardless of the quality of their connection.
However, fairness is hard to crack since it has a logarithmic growth as the number of hard-to-prepare entangled resource states increases. Still, since a multi-client switch with a fair client resource distribution, i.e., a switch with close-in-quality users, is intrinsically fair, the simulation suggests that a switch should establish close-in-quality connections in a single protocol round. 

We suggest two possible directions to further this work. 
Firstly, in describing the optimal allocation in the data center switch network, we mentioned how allocations can be identified towards connections instead of clients, and they turn out to be one and the same in this case as the clients identify connections. 
In more general multi-client scenarios, connection-based allocation is in fact crucial to ensure that all connections are served. 
A different link matching protocol heuristic could be implemented at the switch, where after assigning the required number of resources to each connection, links are ranked at clients locally within each connection they are a part of. 
Clearly such a local-matching protocol would favor fairness of rates across active connections, but each individual end-to-end rate and naturally the overall switch rate be decreased compared to those achieved by the GERM protocol. 
The local matching heuristic may be studied thoroughly in future works and be of interest to networks where fairness is the number one priority.
Another direction for future work is the optimal placement of the switch within an area of an $M$-node, $V$-edge graph of network nodes. 
To start this investigation, we need an easy-to-compute, close expression for the optimum discard window size for arbitrary inter-nodal spacings, which can be established empirically numerically.


\begin{figure*}
\centering
\includegraphics[width=0.8\textwidth]{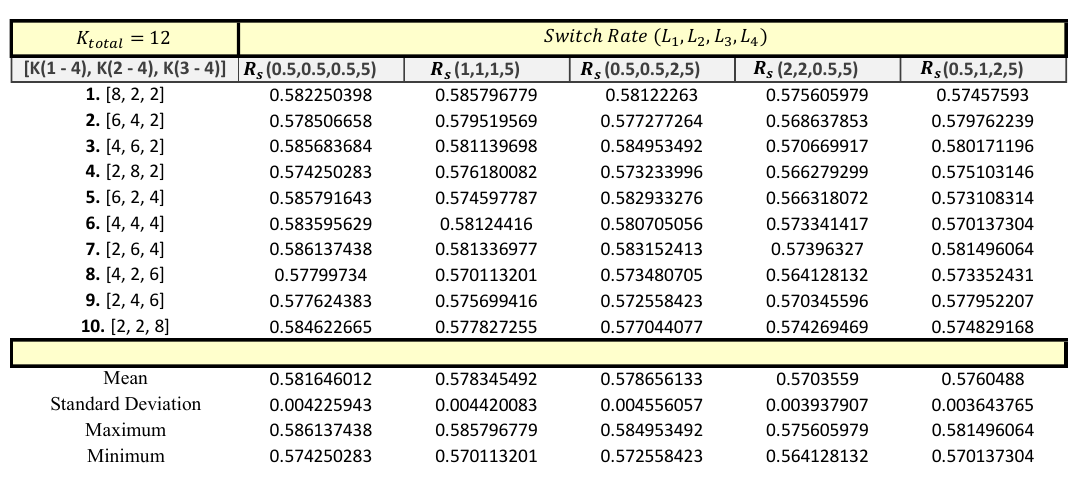}
\includegraphics[width=0.8\textwidth]{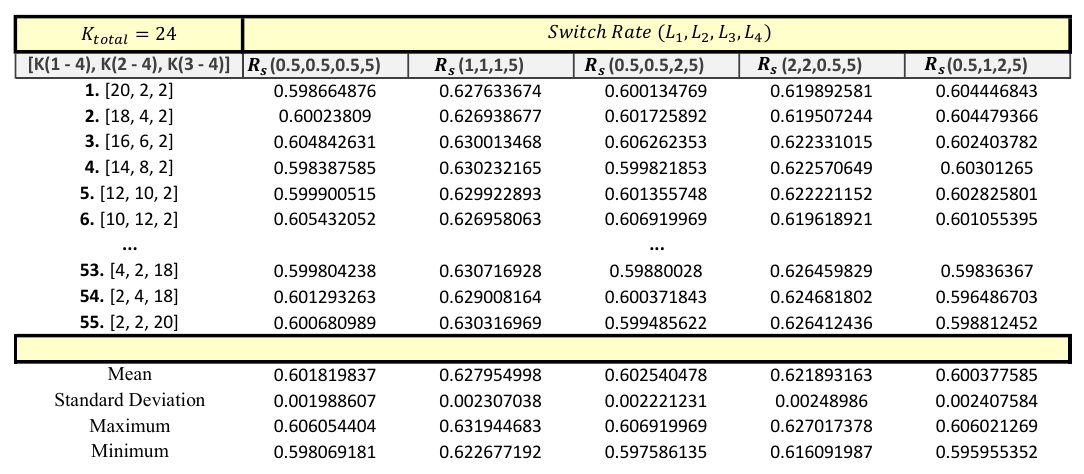}
\includegraphics[width=0.8\textwidth]{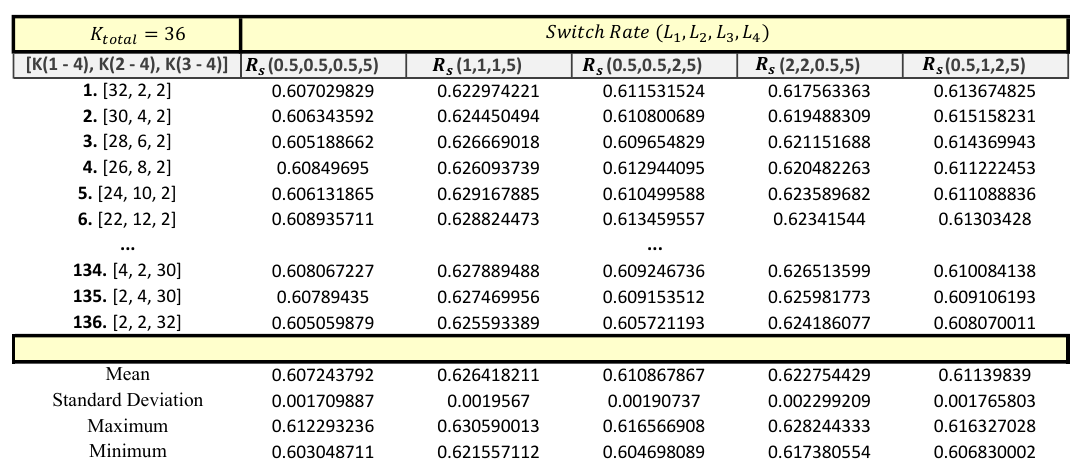}
\caption{
Comparison of total switch rates $R_{\textrm{s}}$ across all the possible resource allocations $[k(1\leftrightarrow 4), k(2\leftrightarrow 4), k(3\leftrightarrow 4)]$ between different connections for a switch network with a data center and three other clients, with a total number of resource states $k_{\textrm{total}} = \sum_{i=1}^{3} k(i\leftrightarrow 4)$ available at the switch. 
Each table considered a variety of clients' settings $(l_1, l_2, l_3, l_4)$. 
As the total number of resource states ($k_{\textrm{total}}$) increase, the overall switch rate ($R_{\textrm{s}}$) grows as well. 
The identical feature between these tables is the $R_{\textrm{s}}$ independence from clients' allocation. 
The reason is behind how we defined a data center or a user ($l_i \ll l_n$, $l_i < 3\textrm{km}$) so that the data center controls the overall switch logical error.
}
\label{fig:multi_client_table}
\end{figure*}

\begin{appendices}


\section{RESOURCE STATE GENERATION AND DISCARD WINDOW}
\label{sec:App_1}

Every quantum helper node (a quantum repeater or more generally a quantum switch) connects clients by generating elementary entanglement links using entangled resource states followed by entanglement swapping. 
This paper follows a multiplexed architecture, where overall $k_{\textrm{total}}$ entangled resource states are generated at each helper node, e.g., see Fig.~\ref{fig:asy_two_cli}. 
Each resource state is a Bell pair between one physical GKP state and one logical GKP state encoded in a [[7,1,3]] Steane code. 
In total, each resource state consists of 8 physical GKP states. 
Each helper node station keeps the [[7,1,3]] logical GKP qubit part of the Bell state referred as inner leaf qubits and sends remaining physical GKP qubit, which is the other part of the Bell state toward the other party as outer leaves. 
This choice is because the inner leaf qubits go through a fiber two times longer than the outer leaf qubits. 
So, an additional error-correcting layer (GKP + Steane code) better protects stored quantum information from noise and photon loss. 
We will not go through the procedure of resource state generation; one can examine~\cite{RSPJG23} for a detailed discussion on the entangled eight GKP resource state cube. 
However, below, we discuss why we only use specific inter-repeater spacing $L \in \{0.5,1,2,2.5,5\}$. 

Increasing the discard window $\nu$ suppress the errors at the expense of requiring a large number of primer resources to generate a required number of the entangled resource state. 
We need to increase the discard window just so much so that the errors from resource state generation are at the order of magnitude of the inner leaf qubit storage errors and communication channel errors. 
The error from resource generation being any better does not help anyway. 
This efficient value for the discard window is related to the inter-nodal spacing $L$ across which the entangled resource state is used for elementary entanglement generation. 
For a large/small inter-nodal spacing, we face a large/small communication channel error, and so we need the inner leaf qubit storage errors to be at large/small. 
This means the optimal discard window should be small/large. 

For values of $L \in \{0.5,1,2,2.5,5\}$km, reasonable choices for the corresponding discard window sizes $\nu$ were found to be $\{7,6,5,4,3\} \times \sqrt{\pi}/{20}$, respectively~\cite{RSPJG23}, by numerical investigation. 
Optimal values of $\nu$ can be similarly determined for other values of $L$.

\section{RESOURCE ALLOCATION AND SWITCH RATE}
\label{sec: App_2}

In this appendix, we discuss the case where one of the clients of a switch is dominant, and how in such a case 
changing the other clients' resource allocation scarcely affects the overall switch rate $R_{\textrm{s}}$. 
We numerically simulate a switch with a total number of entangled resource states $k_{\textrm{total}}$, connecting $n\text{ (other)\ clients}$ to $1\text{ data center}$ in each protocol round. 
Acceptable connections are ones from any of the other clients to the data center. 
Accordingly, half of the resource states are assigned to the data center ($k_{\textrm{total}}/2$). 
The question is, how will the total switch rate ($R_{\textrm{s}}$) be affected by changing the allocation of the remaining resource states ($k_{\textrm{total}}/2$) shared between the users? 
Figure~\ref{fig:multi_client_table} tries to address this question over three story lines. 

The table in Fig.~\ref{fig:multi_client_table} shows the total switch rate $R_{\textrm{s}}$ connecting three clients indexed as $\{1,2,3\}$ and a data center $\{4\}$ over a variety of distances (distance values and $k_{\textrm{total}}$ are assumed to be known beforehand). 
The allowable connections are: (client-1 $\leftrightarrow$ data center), (client-2 $\leftrightarrow$ data center) and (client-3 $\leftrightarrow$ data center). 
The first three columns of the table represent every possible resource allocation, where allocations are shown per connection instead of per client. 
For example, $k(i\leftrightarrow 4)$ is the total number of entangled resource states allocated to the connection of client-$i$ and the data center such that $\sum_{i=1}^{3} k(i\leftrightarrow 4) =k_{\textrm{total}}$.
Upon examining the table, we find that for a choice of ($l_1, l_2, l_3, l_4$) across the column, through all the possible allocations of the $k_{\textrm{total}}$ resources (sharing between the connections), the values of $R_{\textrm{s}}$ are concentrated around the mean. 
Standard deviations are found to be $\ll 0.01$, with the maximum and minimum being close to the average. 
Since the data center is the dominant client and $l_i \in \{0.5,1,2\}\textrm{km}, \text{for } i \in \{1,2,3\}$, the switch sees all the other clients as a single client, requesting to connect to the data center.

Not all switches have a dominant client (clients can be in the same order), but if they do, from the switch perspective, the dominant client governs the error. 
Let's study this from a simpler view. Two clients client-1, client-2, with resource states shared equally between them ($k_1 = k_2 = 10$), are connecting via the proposed repeater in Fig.~\ref{fig:asy_two_cli}. 
We simulate two sets of distances, namely, $\{(l_1=5, l_2=5)$km and $(l_1=0.5, l_2=5)\}$km. 
The results are: $R_{\textrm{s}}(l_1=5, l_2=5)$km $\approx R_{\textrm{s}}(l_1=0.5, l_2=5$km. This means that when one of the two clients controls errors, there is no benefit in improving the other client. 
That's why assigning most resources to the $\text{client-}i\leftrightarrow$ data center connection where $\text{client-}i$ is the closest client to the switch will not favorably increase the switch rate. 
In our analysis, we assign each connection at least two links so assigning all the resources $k_{\textrm{total}}$ to a single connection is prohibited.  

\end{appendices}

\bibliographystyle{IEEEtran}
\bibliography{library.bib}

\end{document}